\documentclass{PoS}
\def\beq{\begin{equation}}
\def\eeq{\end{equation}}
\def\bea{\begin{eqnarray}}
\def\eea{\end{eqnarray}}
\newcommand{\bear}{\begin{eqnarray}}
\newcommand{\eear}{\end{eqnarray}}

\title{Dynamical Electroweak Symmetry Breaking at the LHC on Supercomputers and in the Sky (Asymmetric Dark Matter) }

\ShortTitle{Technicolor going strong.}

 \author{\speaker{Francesco SANNINO}
 \\
    Center for High Energy Physics, University of Southern Denmark, Odense M, DK-5230, Denmark.\\
    E-mail: \email{sannino@ifk.sdu.dk}}

\abstract{ I briefly review the basic challenges and virtues of models breaking the electroweak symmetry dynamically. I will then  introduce the (ultra) minimal walking technicolor models whose construction has been made possible thanks to recent progress in the understanding of the phase diagram for strongly coupled theories as function of number of flavors, colors and matter representation.  I will mention possible relevant collider signatures. Interestingly, asymmetric Dark Matter is a natural possibility in our models providing interesting candidates for decaying Dark Matter models which have been explored to account for the PAMELA and ATIC excesses in $e^\pm$ cosmic rays.  }

\FullConference{Physics at LHC 2008\\
		 29 September - October 4, 2008\\
		 Split, Croatia}

\begin{document}
Understanding the origin of the electroweak symmetry breaking and its possible relation with Dark Matter (DM) constitute two of the most profound theoretical challenges at present. New strong dynamics at the electroweak scale of the type reviewed in \cite{Sannino:2008ha}, inspired to earlier technicolor (TC) models \cite{Weinberg:1979bn}, may very well provide a solution to the problem of the origin of the bright and dark mass.  The two most recent and phenomenologically relevant models are: Minimal  and Ultra Minimal Walking Technicolor \cite{Sannino:2004qp,Ryttov:2008xe}. Minimal walking models and the phase diagram of strongly coupled theories \cite{Ryttov:2007cx,Dietrich:2006cm} are triggering much work spanning from Beyond SM phenomenology \cite{Foadi:2007ue, Belyaev:2008yj,Christensen:2005cb,Gudnason:2006mk} to lattice studies \cite{Catterall:2007yx} 
and cosmology \cite{Cline:2008hr, Gudnason:2006yj,Nardi:2008ix}.

\vskip .3cm
\noindent
{\bf S-parameter:}
New strong dynamics is needed since the oldest TC models featuring QCD-like dynamics  are at odds with electroweak precision tests  (the S-parameter problem (see \cite{Sannino:2008ha} for a modern review)).
The simplest estimate of the contribution to S yields 
$\displaystyle{S_{TC}\approx N_D \frac{d(R_{\rm TC})}{6\pi}}$, 
with $N_D$ the number of doublets with respect to the weak interactions of techniquarks transforming according to the representation $R_{TC}$ of the TC gauge theory  and $d(R_{\rm TC})$ is the dimension of the techniquark representation. The more TC matter is gauged under the electroweak theory the more positive is the S$_{TC}$ parameter.  The full S can receive important contributions also from other sectors, e.g. a fourth family of Dirac leptons. The contribution of the new sector ($ S_{\rm NS}$) can be large, negative, and in several cases computable and hence (in first approximation)
$S = S_{\rm TC} + S_{\rm NS}$.

 \noindent
 {\it TC models can be constrained, via precision measurements, only model by model and the effects of new sectors must be properly included.}

\vskip .3cm
\noindent
{\bf SM - fermion masses:}
Besides breaking the electroweak symmetry the SM Higgs serves the purpose to provide mass to the SM fermions via operators of the type:
$\displaystyle{-Y_d^{ij}\bar{Q}_L^i H d_R^j  - Y_u^{ij}\bar{Q}_L^i (i\tau_2 H^{\ast}) u_R^j }$  $+{\rm h.c.}$,  
where $Y_{q}$ is the Yukawa coupling constant, $Q_L$ is the quark
left-handed Dirac spinor, $H$ the Higgs doublet and
 $q$ the quark right-handed Weyl spinor and $i,j$ the flavor indices. The $SU_L(2)$ weak and spinor indices are suppressed. TC, per se, cannot provide mass to the SM fermions and different approaches have been tried. The most conservative one assumes that no fundamental scalars exist in Nature.  The Yukawa terms become then four-fermion operators and  are naturally interpreted as low energy operators induced by a new strongly coupled gauge dynamics emerging at energies higher than the electroweak theory. These models have been termed extended TC nteractions (ETC) \cite{Eichten:1979ah}.  ETC  most severe limitations are: i) the poor knowledge of strongly coupled dynamics, ii) the need to suppress flavor changing neutral currents (FCNC)  can lead at too small quark masses. A way to alleviate the FCNC problem and reduce the contribution to the S-parameter is to introduce near conformal (walking) dynamics \cite{Holdom:1981rm}.
\vskip .3cm
\noindent
{\bf Unification of the SM Gauge Couplings:}
It is possible to unify the SM gauge couplings when considering TC extensions of the SM. {}For example it has been shown in \cite{Gudnason:2006mk} that near conformal TC type dynamics helps focusing the SM couplings at high energy. ETC interactions are, however, harder to reconcile within a grand unified scheme \cite{Chen:2008jn}. 

\vskip .2cm
\noindent
{\bf Asymmetric Dark Matter (DM):} The amount of baryons in the Universe $\Omega_B\sim 0.04$ is
determined solely by the cosmic baryon asymmetry $n_B/n_\gamma \sim 6\times
10^{-10}$.  In
contrast, we do not know if the DM density is determined by thermal freeze-out, an asymmetry, or another mechanism. What we do know is that if $\Omega_{\rm DM}$ is determined by thermal freeze-out, its proximity to $\Omega_B$
is just a coincidence. If, however, $\Omega_{\rm DM}\sim \Omega_B$ is not accidental, then may be a result of a common asymmetry.
Such a condition is naturally realized in TC models \cite{Nussinov:1985xr,Ryttov:2008xe,Gudnason:2006yj,Nardi:2008ix}. The DM candidate is identified with the lightest technibaryon. If the DM abundance is due to a cosmic asymmetry it will not annihilate but decay. What it is very interesting is that decaying DM particles of the type naturally present in TC models can explain the 
 PAMELA \cite{Adriani:2008zr} and ATIC \cite{ATIC-2} reported excesses in $e^\pm$ cosmic rays, actually privileging an SU(2) TC gauge theory with Ultra Minimal Technicolor structure in \cite{Ryttov:2008xe} .

\vskip .2cm
\noindent
{\bf  Techni-Unparticle (A Natural Model of Unparticle):} In \cite{Sannino:2008nv} we introduced a framework in which the Higgs and the unparticle sector \cite{Georgi:2007ek} are both composite. We sketched a possible unification of these two
sectors at a scale much higher than the electroweak scale. The resulting model resembles
ETC models and we termed it extended technicolor unparticle (ETU). Due to the fact that the model is technically natural it is  internally consistent forbidding any unnatural ultraviolet or infrared divergence.

\vskip .3cm
\noindent
{\bf \large Minimal TC Models}
\vskip .1cm
To have a very low S-parameter and simultaneously reduce the tension with FCNC one would ideally have a TC theory allowing with only one doublet to break the electroweak symmetry dynamically but at the same time being walking (near conformal (NC)). According to the phase diagram put forward in \cite{Sannino:2008ha,Sannino:2004qp,Ryttov:2007cx,Dietrich:2006cm}  the promising candidate theories with the properties required are either theories with fermions in the adjoint representation or two index symmetric one. The relevant feature, found first in \cite{Sannino:2004qp} is that the symmetric-type theories can be NC already for 2 Dirac flavors for SU(2) and SU(3) TC gauge theory. This
should be contrasted with the case of fermions in the fundamental
representation for which the minimum number of flavors required to
reach the conformal window must be larger than 8 already for SU(2) as predicted first in \cite{Ryttov:2007cx} \footnote{We observe that this bound of the conformal window for fundamental representation is a nontrivial prediction of the all-order beta function which appeared few weeks before being confirmed by recent numerical results \cite{Catterall:2007yx}.}. We refer with {\it minimal} to theories for which the number of flavors needed to achieve an infrared fixed point is very small compared to the case of matter in the fundamental representation of the gauge group. 

\vskip .2cm
{\bf Minimal Walking TC (MWT):} We consider an SU(2) TC gauge theory with two adjoint
technifermions \cite{Sannino:2004qp}. According to the ladder approximation  \cite{Sannino:2004qp} this theory is NC while it can possess an infrared conformal fixed point according to the all-order beta function results \cite{Ryttov:2007cx}. The two adjoint fermions are conveniently written as \beq Q_L^a=\left(\begin{array}{c} U^{a} \\D^{a} \end{array}\right)_L , \qquad U_R^a \
, \quad D_R^a \ ,  \qquad a=1,2,3 \ ,\eeq with $a$ being the adjoint color index of SU(2). The left handed fields are arranged in three
doublets of the SU(2)$_L$ weak interactions in the standard fashion. If the theory is NC the condensate is $\langle \bar{U}U + \bar{D}D \rangle$  correctly breaking the electroweak symmetry.

The model, as described so far, suffers from the Witten topological anomaly \cite{Witten:1982fp}. This can be fixed by
adding a new weakly charged fermionic doublet which is a TC singlet. Schematically: 
\beq L_L =
\left(
\begin{array}{c} N \\ E \end{array} \right)_L , \qquad N_R \ ,~E_R \
. \eeq 
The low-energy effective theory to be tested at the LHC, the comparison with precision data and a first study of the unitarity of $WW$ longitudinal scattering can be found in \cite{Foadi:2007ue}. In \cite{Gudnason:2006mk} we discussed the unification issue within this model. 
\vskip.2cm
\noindent
{\bf Ultra Minimal Walking TC (UMT): }
In \cite{Ryttov:2008xe} we provided an explicit example of walking TC with two types of technifermions, i.e. transforming according to two different representations of the underlying TC gauge group. The model possesses a number of interesting properties: i) Features the lowest possible value of the naive S parameter while possessing a dynamics which is NC; ii) Contains, overall, the lowest possible number of fermions; iii) Yields natural asymmetric DM candidates. We termed this model {\it Ultra Minimal near conformal TC} (UMT). It is constituted by an $SU(2)$ TC gauge group with two Dirac flavors in the fundamental representation also carrying electroweak charges, as well as, two additional Weyl fermions in the adjoint representation but singlets under the SM gauge groups. To arrive at this specific UMT model we used the conjectured all-orders beta function for nonsupersymmetric gauge theories \cite{Ryttov:2007cx}. Several low-energy composite particles are SM singlets. In particular there is a di-techniquark state. This TC Interacting Massive Particle (TIMP) is a natural cold DM candidate \cite{Ryttov:2007cx}. We also estimated the fraction of the mass in the universe constituted by our DM candidate over the baryon one as function of the Lepton number and the DM mass. The new TIMP, differently from earlier models \cite{Nussinov:1985xr}, can be sufficiently light to be directly produced and studied at the LHC. If the TIMP is heavy, of the order of the TeV, it is an interesting candidate for explaining the recently reported excesses in $e^\pm$ cosmic rays \cite{Nardi:2008ix}.

\vskip.2cm
\noindent
{\bf Collider  Phenomenology: } The first comprehensive low energy effective theory for MWT, featuring the degrees of freedom  relevant for collider phenomenology, has appeared in \cite{Foadi:2007ue}. The Lagrangian features a (light) composite Higgs as well as the first excited spin one states, i.e. the axial and the vector ones. In \cite{Belyaev:2008yj} we studied and compared the Drell-Yan (DY) and Vector Boson Fusion (VBF) mechanisms for the production of composite heavy vectors at LHC. We found that the heavy vectors are most easily produced and detected via the DY processes. The composite Higgs phenomenology were also studied and the associate production of the Higgs is also a very interesting signal to  explore. 

\vskip .2cm
\noindent
{\bf  Walking on the Lattice: } MWT and  Next to MWT (NMWT) (an SU(3) theory with fermions in the two-index symmetric) are being investigated on the lattice with preliminary indications that  (as predicted in \cite{Sannino:2004qp,Ryttov:2007cx} ) they might possess a (near) conformal dynamics. 
In \cite{Sannino:2008pz} it was shown that  to uncover the presence of an infrared fixed point on the lattice one can use the generalized Gell-Mann Oakes Renner relation.

\end{document}